\documentclass[conference,10pt]{IEEEtran}
\IEEEoverridecommandlockouts
\usepackage[left=0.625in, right=0.625in, top=0.75in, bottom=1in]{geometry}
\usepackage{cite}
\usepackage{amsmath,amssymb,amsfonts}
\usepackage{graphicx}
\usepackage{textcomp}
\usepackage{xcolor}

\usepackage{caption}
\usepackage{subcaption}
\usepackage{lipsum}
\usepackage{url}
\usepackage{verbatim}
\usepackage{todonotes}

\usepackage{authblk}

\usepackage{color}

\usepackage{graphics}
\usepackage{mathtools}
\usepackage{epstopdf}
\usepackage{bbm}
\usepackage{algorithm}
\usepackage[noend]{algpseudocode}
\usepackage{dsfont}
\usepackage{bm}
\usepackage[left=0.625in, right=0.625in, top=0.75in, bottom=1in]{geometry}

\def\BibTeX{{\rm B\kern-.05em{\sc i\kern-.025em b}\kern-.08em
    T\kern-.1667em\lower.7ex\hbox{E}\kern-.125emX}}
\begin{document}


\title{Fast Federated Edge Learning with \\ Overlapped Communication and Computation and\\ Channel-Aware Fair Client Scheduling
}

\author[$\dag$]{ Mehmet Emre Ozfatura}
\author[*]{Junlin Zhao}
\author[$\dag$]{Deniz G\"und\"uz}
\affil[$\dag$]{Department of Electrical and Electronic Engineering, Imperial College London, London, UK}
\affil[*]{School of Science and Engineering, the Chinese University of Hong Kong (Shenzhen), Shenzhen, China
\authorcr Email:
m.ozfatura@imperial.ac.uk,
zhaojunlin@cuhk.edu.cn, d.gunduz@imperial.ac.uk}

\maketitle

\begin{abstract}
We consider federated edge learning (FEEL) over wireless fading channels taking into account the downlink and uplink channel latencies, and the random computation delays at the clients. We speed up the training process by overlapping the communication with computation. With fountain coded transmission of the global model update, clients receive the global model asynchronously, and start performing local computations right away. Then, we propose a dynamic client scheduling policy, called MRTP, for uploading local model updates to the parameter server (PS), which, at any time, schedules the client with the minimum remaining upload time. However, MRTP can lead to biased participation of clients in the update process, resulting in performance degradation in non-iid data scenarios. To overcome this, we propose two alternative schemes with fairness considerations, termed as age-aware MRTP (A-MRTP), and opportunistically fair MRTP (OF-MRTP).
In A-MRTP, the remaining clients are scheduled according to the ratio between their remaining transmission time and the update age, while in OF-MRTP, the selection mechanism utilizes the long term average channel rate of the clients to further reduce the latency while ensuring fair participation of the clients. It is shown through numerical simulations that OF-MRTP provides significant reduction in latency without sacrificing test accuracy. 
\end{abstract}


\section{Introduction}

Extensive research efforts have been devoted to overcome the communication bottleneck in federated learning (FL) \cite{mcmahan2017communication}.
Lossy compression techniques, including quantization\cite{seide20141,alistarh2016qsgd,wen2017terngrad,zhou2016dorefa} and sparsification\cite{strom2015scalable,7835789,aji2017sparse,wang2018atomo} have been developed to reduce the communication cost.
However, these approaches treat the communication channel connecting the clients to the parameter server (PS) as an error-free bit pipe, ignoring the wireless channel characteristics. However, efficient implementation of FL at the wireless edge, called \textit{federated edge learning (FEEL)}, requires jointly optimizing the learning framework with the underlying communication framework taking into account channel characteristics and constraints \cite{Gunduz2020, chen2021distributed}. 

As common in wireless networks, different clients may have distinct computational capabilities, channel statistics, power budgets, etc., resulting in the \textit{straggler effect} in FEEL. By carefully designing the resource allocation policy, the balance between the training latency and energy consumption has been studied in \cite{zeng2020energy,Yang2019,Chen2020}. By scheduling only a limited number of clients in each round in FEEL, the communication requirement can be reduced, and the straggler effect can also be alleviated \cite{nishio2019client, Yang2020d, Yang2020e, 9337227}. In \cite{nishio2019client}, the authors propose to maximize the number of scheduled clients in each round to speed up training. In \cite{Yang2020d}, a function of the ages of client updates is minimized to schedule all the clients as often as possible. Convergence rates of three different scheduling policies, namely, random, round-robin, proportional fair scheduling, are analyzed in \cite{Yang2020e}. Different from the aforementioned works, \cite{9337227} considers `update-aware' client scheduling in FEEL, which takes the significance of model updates into account together with their channel states. Alternatively, in \cite{HFL} clients are clustered around several access points to perform FL hierarchically in order to reduce the communication latency. Another approach to addressing the communication bottleneck is aggregating the local model updates using the superposition property of the wireless multiple access channel \cite{MohammadiAmiri2020, Zhu2020, Amiri2020c}. However, this approach requires accurate synchronization among the transmitting agents \cite{shao2021federated}.

Unlike the previous literature on FEEL, we consider both the downlink and uplink channel variations, together with random computation delays. With the exception of \cite{amiri2020convergence}, works on FEEL ignore the downlink communication channel and the associated latency. Our first contribution is to overlap the downlink and uplink transmissions with local computations at the clients. This is achieved by fountain-coded delivery of the global model update \cite{Castura:CL:06}, such that clients with better downlink channel conditions can receive the global model quickly, and start computing right away. Then, the clients are scheduled for the uplink delivery of their model updates to the PS as soon as they complete their local computations. To further minimize the latency, we schedule the client with the minimial uplink latency  at any point in time.


While this approach, called MRTP, minimizes the latency, it may result in a loss in test accuracy as clients close to the PS would dominate the training process thanks to their statistically better channel conditions. We mitigate this  bias by introducing both short-term and long-term fairness conditions. In particular, we utilize the {\em age} and {\em frequency} metrics, respectively, for short- and long-term fairness, where age, in a broad sense, measures the staleness of the client model update, and frequency measures the long term participation statistics of the clients. Our numerical results show that the proposed overlapped and fair scheduling policy significantly speeds up FEEL without sacrificing the final test accuracy. 

\section{System Model}

\subsection{FL model}
Consider $K$ clients collaboratively training a model parameter vector of dimension $d$, $\bm{\theta}\in\mathbb{R}^d$, with periodic communication with a PS to minimize the empirical loss function, $F(\bm{\theta}) = 1/K \sum_{k=1}^K F_k(\bm{\theta})$, where $F_k(\bm{\theta})$ is the loss function of client $k$.
Let $\mathcal{D}_k$ denote the dataset at client $k$ with $D_k \triangleq |\mathcal{D}_k|$ samples.
The empirical loss function at client $k$ is $F_k(\bm{\theta}) = 1/D_k \sum_{\bm{u}\in\mathcal{D}_k} f(\bm{\theta},\bm{u})$, where $f(\bm{\theta},\bm{u})$ is the task-dependent loss function measured with the model parameter vector $\bm{\theta}$ and data $\bm{u}$.
At each communication round $n \in \{1,2,\cdots \}$, each participant client performs $\tau$-step stochastic gradient descent (SGD) on its local dataset to minimize the loss function based on the received global model parameter $\bm{\theta}(n)$. At the $m$-th step of the local SGD of participant client $k$ in communication round $n$, the local model is updated as
\begin{align}
    \bm{\theta}_k^{m+1}(n)=\bm{\theta}_k^{m}(n) - \eta_k^m(n) \nabla F_k(\bm{\theta}_k^m(n),\xi_k^m(n)),
\end{align}
where $\eta_k^m(n)$ is the learning rate, and $\nabla F_k(\bm{\theta}_k^m(n),\xi_k^m(n))$ is the gradient computed with the local model parameter $\bm{\theta}_k^m(n)$ and mini-batch data $\xi_k^m(n)$, which is uniformly randomly selected from $\mathcal{D}_k$ and follows $\mathbb{E}_\xi\{\nabla F_k(\bm{\theta}_k^m(n),\xi_k^m(n))\}=\nabla F_k(\bm{\theta}_k^m(n))$. 

Each participant client then forwards its updated local model to the PS. Denoting the model vector at client $k$ updated after $\tau$-steps by $\bm{\theta}_k(n)$, $k=1,\cdots,K$, the global model is updated by the PS as
$\bm{\theta}(n+1) =\frac1K \sum_{k=1}^K \bm{\theta}_k(n)$.


\subsection{Communication Model}
A block fading channel model is assumed, where the channels between the PS and the clients remain unchanged in each communication round of the FEEL process.

\subsubsection{Asynchronous global model transmission}
Note that we have a multicast channel when transmitting the global model from the PS to the clients. To speed up the training process, we use fountain coded multicasting of the global model to the clients \cite{Castura:CL:06}. The downlink rate at client $k$ in communication round $n$ is given by $R_k^{\text{dl}}(n) = \text{log}_2~\left(1+\frac{P\vert h_k^{dl}(n)\vert^2}{\sigma_k^2} \right)$, $\forall k$, where $h_k^{dl}(n)$ is the complex-valued downlink channel coefficient between the PS and client $k$ in round $n$, $P$ is the transmit power at the PS, and $\sigma_k^2$ is the noise variance. 


With asynchronous global model transmission in the downlink, a client recovers the global model with a latency dependent on its channel gain, and immediately starts local computations. Assuming that the global model is compressed into $Q$ bits, it will take $Q/R_k^{\text{dl}}(n)$ seconds for client $k$ to receive the model update. This will allow us to parallelize global model transmission and computations, instead of targeting the worst client to guarantee all the clients receive the global model simultaneously. 




In the rest of the paper, to simplify notation we will drop the round index $n$ when it is clear from the context. 

\subsubsection{Local model update} In the uplink, where clients upload their model parameters to the PS, we assume a time-division framework; that is, only a single client is scheduled at any point in time. The instantaneous rate of each client is given by $R_k^{\text{ul}}(n) = \text{log}_2\left(1+P_k\vert h_k^{ul}(n)\vert^2 / \sigma_0^2 \right),$ where $P_k$ is the transmit power at client $k$, and $h_k^{ul}(n)$ is the uplink channel coefficient from client $k$ to the PS. 

\subsection{Local computations} 
We assume that the computation speeds at the clients are also random following a distribution that is independent across clients and rounds. In each round, we denote by $\pi(j)$ the client with the $j$-th smallest accumulated latency in downlink transmission and local computation, $j=1,\ldots, K$, and by $T_j$ the corresponding latency since the beginning of the communication round, with $T_0 \triangleq 0$.



\subsection{Client scheduling}


With asynchronous downlink transmission and heterogeneous local computation speeds, clients complete their local model updates in a sequential manner. The clients that have completed local computations and are thus available to upload their local models to the PS are referred to as {\em idle} clients. We denote the set of idle clients at time $t$ within each round by $\mathcal{C}^{idle}(t)$.
Hence, client $\pi(j)$ is added to the idle set at time $T_j$, and remains there until it uploads its model to the PS. 



Let $s(t) \in \{0, 1, \ldots, K\}$ denote the index of the client scheduled for transmission at time $t$, where $s(t)=0$ means no client is scheduled. We have $s(t) \neq \pi(j)$ if $t < T_j$; that is, a client cannot be scheduled for upload before it completes its local computations. Let $Q_{k}(t)$ denote the remaining size of model parameter vector (measured in bits) at time $t$ that has not yet been uploaded to the PS. We have $Q_{\pi(k)}(T_k)= Q$, and
\begin{align}\label{eq:bits}
    Q_{k}(t) = Q - R_k^{\text{ul}} \int_{0}^t \mathds{1}_{\{s(t)=k\}} dt,
\end{align}
where $\mathds{1}_{\{x\}}$ is the indicator function, which is 1 when $x$ holds, and $0$ otherwise. 


Let $t_k$ denote the time client $k$ completes uploading its model to the PS, i.e., $t_k \triangleq \min_{t} \{Q_k(t) =0\}$. Client $k$ is removed from the idle set at time $t_k$. Each round continues until $N$ out of $K$ clients upload their models to the PS, and therefore, some clients may never be added to the idle set, or may not leave the set at the end of the round. We have $t_k = \infty$ for those clients. Let $\mathcal{K}(t)$ denote the set of clients that have completed their upload by time $t$ within round $n$, i.e., $\mathcal{K}(t) = \{k: t_k \leq t\}$. For a specific scheduling policy, the completion time of round $n$ is given by  $T(n) \triangleq \min_t \{|\mathcal{K}(t)| = N\}$, while the set of clients scheduled in round $n$ are given, with slight abuse of notation, by $\mathcal{K}_n \triangleq \mathcal{K}(T(n))$.

\section{Scheduling Policies}

Our goal is to come up with scheduling policies that would not only minimize the completion time of each round, but also lead to the fastest convergence in terms of the wall-clock time.

\begin{algorithm}[t]\small
\caption{MRTP in round $n$}\label{mrtp}
\begin{algorithmic}[1]
\Require{$K, N, Q, \{R_k^{\text{ul}}(n)\}_{k=1}^K$}
\Ensure{$\mathcal{K}_n$, $T(n)$}
\State{Initialization: $t=0, \mathcal{C}^{idle}(t)=\emptyset,\mathcal{K}_n =\emptyset, Q_k(t)=Q$}
\While{$\vert\mathcal{K}_n \vert <N$}
    \If{$\mathcal{C}^{idle}(t)\neq\emptyset$}
    \State{Schedule $k^* = \text{arg}~\!\min \frac{Q_k(t)}{R_{k}^\text{ul}}$ for $k \in \mathcal{C}^{idle}(t)$}
            \While {No new arrival and $Q_{k^*}(t) > 0$ }
                \State {Schedule client $k^*$, update $t$, $Q_{k^*}(t)$ as in (\ref{eq:bits})}
            \EndWhile
        \If{new arrival $k$}
        \State{Update $\mathcal{C}^{idle}(t)\leftarrow \mathcal{C}^{idle}(t) \bigcup \{k\}$}
        \Else
        \State{Update $\mathcal{C}^{idle}(t)\leftarrow \mathcal{C}^{idle}(t) \setminus \{k^*\}$}
        \State{Update $\mathcal{K}(t)\leftarrow\mathcal{K}(t)\bigcup
        \{k^*\}$}
        \EndIf
    \Else
    \State{Wait until arrival of another client}
    \If{new arrival $k$}
    \State{$t \leftarrow T_{k}$}
    \State{Update $\mathcal{C}^{idle}(t)\leftarrow \mathcal{C}^{idle}(t) \bigcup \{k\}$}
    \EndIf
    \EndIf
\EndWhile
\end{algorithmic}
\end{algorithm}

\subsection{Minimum remaining time-based policy (MRTP)}

Note that, at any time instant within a round, we can schedule any client from the idle set. However, it is easy to see that there is no loss of optimality making scheduling decisions only when the idle set is updated, i.e., when a new client becomes idle, or one of the idle clients completes uploading its model.

In MRTP, each time the idle set is updated, we schedule the client with the minimum remaining time to upload its update. Specifically, at time $t$, client $k^*$ is scheduled if
\begin{align}
    k^* = \text{arg}~\!\min_{k \in C^{idle}(t)} \frac{Q_k(t)}{R_{k}^\text{ul}}.
\end{align}
Details of the MRTP can be found in Algorithm \ref{mrtp}.


\subsection{Age-aware MRTP (A-MRTP)}

While MRTP minimizes the upload time, clients with statistically better channel qualities are more likely to be scheduled, which results in non-uniform sampling of clients and over-fitting due to excessive use of limited amount of data. This may drastically degrade the performance of FEEL, especially when the data is not i.i.d., which is common in practice. Therefore, to strike a balance between the latency and model accuracy, we propose two alternative schemes by taking the `age of update' into consideration. 

For the short term fairness, we utilize the {\em age} metric, where the age of a client at round $n$, denoted by $a_{k,n}$, represents the number of rounds since the last time it was scheduled. The  age parameter evolves as follows:
\begin{equation}\label{age_update}
    a_{k,n+1} =
    \begin{cases*}
      a_{k,n}+1, & if $k\notin \mathcal{K}_n$ \\
      1,        & if $k\in \mathcal{K}_n$
    \end{cases*}.
  \end{equation}

In A-MRTP, $\alpha N$ clients are firstly selected to upload their models as in MRTP to minimize the latency, where $\alpha \in [0,1]$ is a tuning parameter. Then, to promote selecting clients that are less frequently scheduled, we select the client with the minimum ratio between the remaining time and the age of its update, which can be considered as the average latency for each timely update in the short term. Therefore, a balance between the efficiency in training and fairness is achieved by tuning $\alpha$.

\subsection{Opportunistic Fair MRTP (OF-MRTP)}

The main drawback of A-MRTP is that it utilizes only instantaneous rate  $R_{k}^\text{ul}(n)$ for client scheduling. However, when the clients are located at different distances from the PS, their participation frequency will still depend on their locations. 
We propose an opportunistic policy that utilizes the relative channel condition, denoted by $\gamma_{k, n}$, which measures the ratio of instantaneous channel rate of client $k$ to its long term average value, $\bar{R}_{k}^\text{ul}$; that is, we have $\gamma_{k,n}= R_{k}^\text{ul}(n) / \bar{R}_{k}^\text{ul}$. Hence, instead of scheduling clients based on their instantaneous channel states, we use $\gamma_{k,n}$ for scheduling. Further, we consider two metrics to ensure both short-term and long-term fairness among the clients. We use the \textit{age metric} for short-term fairness, that is, to promote uniform participation of the clients, and define a \textit{frequency metric} for long-term fairness, that is, to promote equal participation of clients. The frequency metric, $f_{k,n}$, denotes the participation frequency of the $k$th client at round $n$. We define $f_{k,n} \triangleq l_{k}(n)/(n-1)$, where $l_{k}(n)$ denotes the total number of rounds that client $k$ has participated until round $n$.

In order to introduce long-term fairness, we consider a subset of the idle clients as follows
\begin{equation}\label{eq:OF_set}
\tilde{\mathcal{C}}^{idle}(t) = \{k: k\in \mathcal{C}^{idle}(t), f_{k,n}<f_{max} \},
\end{equation}
where $f_{max}$ is a maximum frequency constraint. 
For the opportunistic policy, we further consider the following subset of clients
\begin{equation}
\hat{\mathcal{C}}^{idle}(t) \triangleq \{k: k\in \tilde{\mathcal{C}}^{idle}(t) ~ ,  ~ a_{k,n}>a_{th} ~ , ~\gamma_{k,n}>\gamma_{min} \},
\end{equation}
where $a_{th}$ is the minimum age constraint introduced to ensure short-term fairness and $\gamma_{min}$ is the rate constraint for opportunistic scheduling. If there are multiple clients in $\hat{\mathcal{C}}^{idle}(t)$, then client $k$ with the  maximum $\gamma_{k,n}$ value, that is  the one with the best relative channel condition is scheduled.

Similarly to A-MRTP, the proposed opportunistic policy consists of two steps. In the initial step, $\alpha$ portion of the clients are scheduled from $\tilde{\mathcal{C}}^{idle}(t)$ according to MRTP, while the remaining clients are scheduled from $\hat{\mathcal{C}}^{idle}(t)$ based on $\gamma_{k,n}$. However, if $\hat{\mathcal{C}}^{idle}(t)=\emptyset$, then client is scheduled according to MRTP from $\tilde{\mathcal{C}}^{idle}(t)$.
As shown in (\ref{eq:OF_set}), clients with excessive participation frequency are excluded from scheduling, which increases the participation of less frequently selected clients, and thus, improves the fairness in scheduling.
Overall, the proposed opportunistic scheduling strategy OF-MRTP is defined by four system parameters $\alpha$, $a_{th}$, $\gamma_{min}$, and $f_{max}$.

\begin{table*}[t]
\begin{center}
\begin{tabular}{ |c | c| c| c|c|}
    \hline
    \textbf{Age threshold } & \textbf{MRTP fraction } & \textbf{frequency constraint } & \textbf{ Accuracy (mean $\pm$ std)} & \textbf{Average per round latency}\\ 
    \hline
    $a_{th}=10$ & $\alpha=0.25$ & $f_{max}=0.3$ & $83.857 \pm0.43$ & $84.81\pm3.41$ seconds\\
    \hline
     $a_{th}=10$ & $\alpha=0.25$ & $f_{max}=0.4$ & $83.405 \pm0.29$ &  $86.6 \pm3.13$ seconds\\
     \hline
      $a_{th}=10$ & $\alpha=0.5$ & $f_{max}=0.3$ & $83.65 \pm0.49$ & $84.49 \pm3.78$ seconds\\
      \hline
       $a_{th}=10$ & $\alpha=0.5$ & $f_{max}=0.4$ & $83.157 \pm0.61$ & $86.33 \pm4.77$ seconds\\
       \hline
       $a_{th}=5$ & $\alpha=0.25$ & $f_{max}=0.3$ & $84.10 \pm0.43$ & $141.35 \pm9.27$ seconds\\
       \hline
     $a_{th}=5$ & $\alpha=0.25$ & $f_{max}=0.4$ & $83.734 \pm0.4$ & $138.05 \pm7.77$ seconds\\
      \hline
      $a_{th}=5$ & $\alpha=0.5$ & $f_{max}=0.3$ & $83.84 \pm0.39$  & $87.25  \pm5.65$ seconds\\
      \hline
       $a_{th}=5$ & $\alpha=0.5$ & $f_{max}=0.4$ & $84.02 \pm0.3$ & $70.04 \pm4.17$ seconds\\
    \hline
\end{tabular}
\end{center}
 \caption{Test accuracy and average per round latency of OF-MRTP for 5000 rounds, averaged over 10 trials. 
 }
 \label{tab:OF_MRTP}
\end{table*}


\begin{table}[b]
\begin{center}
\begin{tabular}{ |c | c| c| c|}
    \hline
    $\textbf{MRTP fraction}$ & \textbf{ Accuracy (mean $\pm$ std)} & \textbf{Average latency}\\ 
    \hline
    $\alpha=0.9$ & $80.67 \pm0.73$ & $315.14\pm151.4$ seconds\\
    \hline
    $\alpha=0.7$ & $82.01 \pm0.375$ & $414.5 \pm164.6$ seconds\\
      \hline
\end{tabular}
\end{center}
 \caption{Test accuracy and average per round latency of A-MRTP for 5000 rounds, averaged over 10 trials. 
 }
 \label{tab:age}
\end{table}

\section{Numerical Results}
\subsection{Simulation Setup}
\subsubsection{Objective and network setup}
We consider image classification on the CIFAR-10 dataset, which contains 50,000 training and 10,000 test images from 10 classes. We employ a convolutional neural network (CNN) architecture consisting of 4 convolutional layers followed by 4 fully connected layers. We set $\tau=4$ local iterations using batchsize of 32. 

We consider $100$ client devices randomly distributed around the PS, and assume that the training dataset is distributed disjointly among the client devices in a non-iid manner, such that each client has 500 distinct training images from at most 4 different classes. Finally, we set the participation ratio to $20\%$, which means, at each round, $N=20$ devices, out of 100, are scheduled to upload their model to the PS.

\subsubsection{Computation latency}
To model the computation latency at the clients, we consider the commonly employed shifted exponential distribution \cite{CC.2}, where the probability of completing $\tau$ local updates by time $t$ is given by
\begin{equation}\label{dist}
I(t)=
    \begin{cases}
     1-e^{-\mu(\frac{t}{\tau}-T_{min})}, &  \text{if } t\geq \tau T_{min}, \\
      0,   &  \text{otherwise}. 
    \end{cases}
\end{equation}
where $T_{min}$ is the minimum computation latency to perform one local update and $\mu$ is the average delay for one local update that is $\mu=\bar{T}-T_{min}$, where $\bar{T}$ is the mean computation time for one local update. 
In order to obtain practically relevant values for $\mu$ and $T_{min}$, we measured the time for computation on a CPU using time.time() command, according to given batchsize and the network model, for over 100000 trials.

\subsubsection{Path loss and noise}
$K$ clients are uniformly randomly distributed in the region within 500 meters around the PS. The path loss of client $k$ is given by $\text{PL}_k = 148.1 + 37.6\text{log}_{10} d_k$, where $d_k$ is the distance of $k$ to the PS measured in kilometers.
We set $P = 15~\text{dBm}$, and $P_k = 10~\text{dBm}, \forall k$.
The channels of clients across different time slots are modeled as i.i.d. fading. The standard deviation of the channel gain of client $k$ is
$\phi_k=(10^{{-\text{PL}_k}/10})^{1\slash 2}$, $\forall k$, and the channel coefficient is modeled as $h_k(n)= \phi_k\tilde{h}_k(n)$, where $\tilde{h}_k(n)$ denotes an i.i.d. random variable accounting for Rayleigh fading of unit power in communication round $n$. The variances of noise at the PS and the clients are set as $\sigma_0^2 =\sigma_k^2=7.96 \times 10^{-14}$ Watts.

\subsection{Simulation Results}
We first consider the A-MRTP scheme. For the experiments, we consider $\alpha= \{0.7, 0.9\}$. The average per round latency and test accuracy results are presented in Table \ref{tab:age}. As predicted, we observe that by decreasing $\alpha$; that is, scheduling more clients according to the age-metric, both the test accuracy and the average latency increase.

We then consider OF-MRTP by setting $\alpha= \left\{0.25,0.5\right\}$, $a_{th}= \left\{5, 10\right\}$, $f_{max}=\left\{0.3, 0.4\right\}$, and  $\gamma_{min}=1$. The  test accuracy and average  per round latency results are presented in  Table \ref{tab:OF_MRTP}. We note that, if a round-robin scheduler is employed with participation ratio $20\%$, then the maximum age will be $a_{t,k}=5$, and similarly, the maximum participation frequency will be $f_{k,t}=0.2$. Therefore, to setup the parameter values  $a_{th}$ and $f_{max}$, we consider these values as our reference points. One can easily observe from Table \ref{tab:OF_MRTP} that with OF-MRTP the average latency as well as the variation on the latency significantly drops, since client $k$ is scheduled only if $R_{k}^\text{ul}(n)>\bar{R}_{k}^\text{ul}(n)$. Besides, thanks to the control on the participation frequency of the clients, we also observe a significant improvement in the test accuracy. We also want to remark that when $a_{th}=10$, we observe similar test accuracy and latency results for other parameters. The reason is that since MRTP is used for client scheduling when $\hat{\mathcal{C}}^{idle}(t)=\emptyset$, at the same time  larger $a_{th}$ increases the probability of $\hat{\mathcal{C}}^{idle}(t)$ being empty, thus more clients are scheduled according to MRTP. This observation is also backed by the simulation results with $a_{th}=5$, where the impact of the parameter $\alpha$ is more visible. Not surprisingly the minimum latency is achieved when more clients are scheduled according to MRTP and we allow larger participation frequencies. Interestingly, in this case, where $\alpha=0.5$ and $f_{max}=0.4$, we do not observe a compromise on the test accuracy.

\begin{figure}[t]
\centering
\includegraphics[scale =0.5]{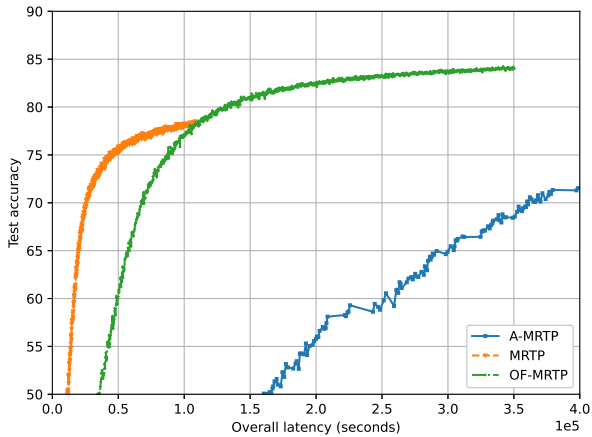}
\caption{Comparison of the MRTP, A-MRTP and OF-MRTP}
\label{comp}
\end{figure}

For comparison, we consider MRTP and random scheduling as benchmark. Note that these are the optimal strategies, respectively, from the latency and fairness perspectives. In Fig. \ref{comp}, we compare the convergence behaviour of the OF-MRTP, A-MRTP and MRTP schemes. As one would expect, the test accuracy increases faster with MRTP; however, due to the non-iid distribution of the data it converges to a sub-optimal model and eventually diverges\footnote{The test accuracy is plotted until the divergence point.}. While A-MRTP eventually reaches an accuray level above MRTP (see Table \ref{tab:age}), it introduces significant latency due to scheduling the clients with poor channel conditions to reduce their age. The convergence behaviour of random scheduling is not included in the figure since the average per-round latency is very high, instead, we compare OF-MRTP and random scheduling based on the final test accuracy results which is averaged over 10 trails. We observe that the average final test accuracy with random scheduling is $83.92 \pm 0.37$. In comparison with the test accuracy results in Table \ref{tab:OF_MRTP}, we can conclude that OF-MRTP does not compromise the accuracy while significantly reducing the latency.

\section{Conclusion}
We proposed novel global model transmission and client scheduling techniques to speed up wall-clock training time for FEEL without sacrificing final test accuracy. In particular, we ensured fair participation of the clients to achieve high test accuracy, and reduced the overall latency, which  includes the computation time and model uplink/downlink latencies. To this end, we first introduced a fountain coded asynchronous model downlink strategy to allow clients to start local computations without waiting for others to download the global model. We then introduced MRTP, which adaptively schedules the client that can upload its local model to the PS in the fastest manner. MRTP and asynchronous downlink strategy, together, help to overlap computation and communication time, thus reduce the overall latency. However, as we experimentally show, client selection that solely focuses on the latency may lead to divergence when certain clients participate in the model update more frequently than others. Hence, we further employed the 'update age' and `update frequency' as fairness metrics, which are opportunistically used to speed up training without sacrificing accuracy. Finally, through extensive simulations we show that it is possible to significantly reduce the overall latency without compromising the test accuracy.


%

\bibliographystyle{IEEEtran}
\bibliography{IEEEabrv,refs.bib}

\ifCLASSOPTIONcaptionsoff
  \newpage
\fi



%

\end{document}